\documentclass[twocolumn]{aastex631}
\usepackage{mathrsfs}
\usepackage{graphicx}
\usepackage{subfigure}
\usepackage{epstopdf}
\usepackage{amsmath}
\usepackage{amssymb}
\usepackage{hyperref}
\usepackage{color}

\begin{document}
\title{Exploring primordial curvature perturbation on small scales with the lensing effect of fast radio bursts}

\author{Huan Zhou}
\affiliation{Department of Astronomy, School of Physics and Technology, Wuhan University, Wuhan 430072, China}

\author{Zhengxiang Li}
\affiliation{Department of Astronomy, Beijing Normal University, Beijing 100875, China}
\affiliation{Institute for Frontiers in Astronomy and Astrophysics, Beijing Normal University, Beijing
102206, China}
\email{zxli918@bnu.edu.cn}

\author{Zong-Hong Zhu}
\affiliation{Department of Astronomy, School of Physics and Technology, Wuhan University, Wuhan 430072, China}
\affiliation{Department of Astronomy, Beijing Normal University, Beijing 100875, China}
\email{zhuzh@whu.edu.cn}

\begin{abstract}
Cosmological observations, e.g., cosmic  microwave background, have precisely measured the spectrum of primordial curvature perturbation on larger scales, but smaller scales are still poorly constrained. Since primordial black holes (PBHs) could form in the very early Universe through the gravitational collapse of primordial density perturbations, constrains on the PBH could encodes much information on primordial fluctuations. In this work, we first derive a simple formula for lensing effect to apply PBH constraints with the monochromatic mass distribution to an extended mass distribution. Then, we investigate the latest fast radio burst observations with this relationship to constrain two kinds of primordial curvature perturbation models on the small scales. It suggests that, from the null search result of lensed fast radio burst in currently available observations, the amplitude of primordial curvature perturbation should be less than $8\times 10^{-2}$ at the scale region of $10^5-10^6~\rm Mpc^{-1}$. This corresponds to an interesting mass range relating to binary black holes detected by LIGO-Virgo-KAGRA and future Einstein Telescope or Cosmic Explorer.
\end{abstract}

\keywords{Primordial black holes, Gravitational lensing, Fast radio bursts.}

\section{Introduction}\label{sec1}
The power spectrum of primordial curvature perturbations on large scales has been precisely constrained by a variety of observations. For instance, cosmic microwave background (CMB) and large scale structure (LSS) observations suggest that the amplitude of primordial curvature perturbation should be orders of magnitude larger than $\mathcal{O}(10^{-9})$ at $\mathcal{O}(10^{-4}-10^0)~\rm Mpc^{-1}$ scales~\citep{CMB2018}. However, most current available cosmological observations are only able to constrain primordial fluctuations at larger scales. Therefore, new probes are in great request to constrain primordial perturbations at smaller scales. Moreover, primordial black hole (PBH) have been a field of great astrophysical interest because they are often considered to make up a part of dark matter. PBH could form in the early Universe through the gravitational collapse of primordial density perturbations~\citep{Hawking1971,Carr1974,Carr1975}, its formation is closely related to the primordial power spectrum~\citep{Sasaki2018, Green2021}. Therefore, there are many inflation models, e.g., inflation model with modified gravity~\citep{Pi2018,Fu2019}, multi-field inflation model~\citep{Clesse2015,Cai2019}, special single-field inflation model~\citep{Cai2020,Motohashi2020}, to enhance the amplitude of power spectrum of primordial curvature perturbations on small scales which corresponds to PBHs in various mass windows.

Theoretically, the mass of PBHs can range from the Planck mass ($10^{-5}~\rm g$) to the level of the supermassive black hole in the center of the galaxy. So far, numerous methods, including both direct observational constraints and indirect ones, have been proposed to constrain the abundance of PBHs in various mass windows~\citep{Sasaki2018, Green2021}. Gravitational lensing effect is one of direct observational probes to constrain the abundance of PBH over a wide mass range from $\mathcal{O}(10^{-10}~M_{\odot})$ to $\mathcal{O}(10^{10}~M_{\odot})$. In general, we can divide the method of lensing effect into four types~\citep{Liao2022}: 1. Searching the luminosity variation of persistent sources~\citep{Allsman2001,Griest2013,Niikura2019a, Niikura2019b,Tisserand2007,Zumalacarregui2018}, for example, observing a large number of stars and looking for amplifications in their brightness caused by lensing effect of intervening massive objects could yield constraints on the abundance of deflectors~\citep{Allsman2001,Tisserand2007,Griest2013,Niikura2019a, Niikura2019b}; 2. Searching multiple peaks structures of transient sources~\citep{Munoz2016,Laha2020,Liao2021,Zhou2022a,Zhou2022,Oguri2022,Krochek2022,Connor2023, Blaes1992, Nemiroff2001,Ji2018, Lin2022}, such as searching echoes due to the milli-lensing effect of fast radio bursts (FRBs) were proposed to put constraints on the PBH abundance~\citep{Munoz2016, Laha2020,Liao2021,Zhou2022a,Zhou2022,Oguri2022,Krochek2022,Connor2023,Kalita2023}; 3. Searching multiple images produced by milli-lensing of possible persistent sources like the compact radio sources (CRSs) can be used to constrain the supermassive PBH~\citep{Press1973,Kassiola1991,Wilkinson2001,Zhou2022c,Casadio2021}; 4. Searching the waveform distortion caused by the lensing effect of distant sources~\citep{Jung2019,Liao2020a,Urrutia2021,Wang2021,Basak2022,Zhou2022d,LIGO2023,Urrutia2023,CHIME2022,Leung2022,Barnacka2012}, for example,  distorting the GW waveform as the fringes were proposed to constrain PBH with stellar mass~\citep{Jung2019,Liao2020a,Urrutia2021,Wang2021,Basak2022,Zhou2022d,LIGO2023,Urrutia2023}. In this work, based on a relationship for applying constraints with the monochromatic mass distribution (MMD) to a specific extended mass distribution (EMD), we proposed to use the lensing effect of fast radio bursts to study the primordial curvature perturbations on small scales which have not been achieved by other observations.

This paper is organized as follows: Firstly, we introduce formation of PBHs from the primordial curvature perturbation model in Section~\ref{sec2},. In Section~\ref{sec3}, we carefully analyzed constraints on PBHs from the lensing effect. In Section~\ref{sec4}, we present the results of constraints on power spectrum. Finally, we present discussion in Section~\ref{sec5}. Throughout, we use the concordance $\Lambda$CDM cosmology with the best-fit parameters from the recent Planck observations~\citep{Planck2018}.

\section{Formation of primordial black holes}\label{sec2}
The power spectrum of primordial curvature perturbations determines the probability of PBH production, the mass function of PBHs, and the PBH abundance~\citep{Sasaki2018,Green2021}. The phenomena of critical collapse could describe the formation of PBHs with mass $m_{\rm PBH}$ in the early Universe, depending on the horizon mass $m_{\rm H}$ and the amplitude of density fluctuations $\delta$~\citep{Carr2016}:
\begin{equation}\label{eq2-1}
m_{\rm PBH}=Km_{\rm H}(\delta-\delta_{\rm th})^{\gamma},
\end{equation}
where $K=3.3$, $\gamma=0.36$, and $\delta_{\rm th}=0.41$~\citep{Harada2013}, and the horizon mass $m_{\rm H}$ is related to the horizon scale $k$~\citep{Nakama2017}
\begin{equation}\label{eq2-1f}
m_{\rm H}\approx17\bigg(\frac{g_{*}}{10.75}\bigg)^{-1/6}\bigg(\frac{k}{10^6~{\rm Mpc^{-1}}}\bigg)^{-2}~M_{\odot}
\end{equation}
where $g_{*}$ is the number of relativistic degrees of freedom. The coarse-grained density perturbation is given by
\begin{equation}\label{eq2-2}
\begin{split}
\sigma^2(k)=\int d\ln q \frac{16}{81} \bigg(\frac{q}{k}\bigg)^4W^2(q/k) T^2(q,k)\times\\
P_{\zeta}(q,\boldsymbol p_{\rm mf}),
\end{split}
\end{equation}
where $W(q/k)$ is the Gaussian window function, $P_{\zeta}(q,\boldsymbol p_{\rm mf})$ is the power spectrum of primordial curvature perturbation, and $T(q,k)$ is the transfer function~\citep{Young2014,Ando2018}
\begin{equation}\label{eq2-3}
T(q,k)=\frac{3(\sin x-x\cos x)}{x^3},
\end{equation}
where $x=\frac{q}{\sqrt{3}k}$. To convert $\sigma^2(k)$ to the mass function of PBHs, we calculate the probability of PBH production by considering the Press-Schechter formalism~\citep{Press1974}
\begin{equation}\label{eq2-4}
\begin{split}
\beta_{m_{\rm H}}=\int_{\delta_{\rm th}}^{+\infty}\frac{m_{\rm PBH}}{m_{\rm H}}P_{m_{\rm H}}(m_{\rm H})d \delta(m_{\rm H})=\\
\int_{-\infty}^{+\infty}\frac{m_{\rm PBH}}{m_{\rm H}}P_{m_{\rm H}}(m_{\rm H})\frac{d\delta(m_{\rm PBH})}{d \ln m_{\rm PBH}} d \ln m_{\rm PBH}=\\
\int_{-\infty}^{+\infty} \bar{\beta}_{m_{\rm PBH}} d \ln m_{\rm PBH},
\end{split}
\end{equation}
where $P_{m_{\rm H}}(m_{\rm H})$ denotes a Gaussian probability distribution of primordial density perturbations at the given horizon scale,
\begin{equation}\label{eq2-5}
\begin{split}
P_{m_{\rm H}}(\delta(m_{\rm PBH}))=\frac{1}{\sqrt{2\pi\sigma^2(k(m_{\rm H}))}}\times\\
\exp{\bigg(-\frac{\delta^2(m_{\rm PBH})}{2\sigma^2(k(m_{\rm H}))}\bigg)}.
\end{split}
\end{equation}
The PBH energy fraction is calculated from Eq.~(\ref{eq2-4}) as
\begin{equation}\label{eq2-6}
\Omega_{\rm PBH}=\int_{-\infty}^{+\infty} d\ln m_{\rm H}\bigg(\frac{M_{\rm eq}}{m_{\rm H}}\bigg)^{1/2}\beta_{m_{\rm H}},
\end{equation}
where $M_{\rm eq}=2.8\times10^{17}~M_{\odot}$ is the horizon mass at the time of matter-radiation equality ~\citep{Nakama2017}. In addition, the mass function of PBHs $\psi(m_{\rm PBH}, \boldsymbol p_{\rm mf})$ can be obtained by differentiating $\Omega_{\rm PBH}$ with the PBH mass
\begin{equation}\label{eq2-7}
\begin{split}
\psi(m_{\rm PBH}, \boldsymbol p_{\rm mf})=\frac{1}{\Omega_{\rm PBH}}\frac{d \Omega_{\rm PBH}}{d m_{\rm PBH}}=
\frac{1}{m_{\rm PBH}\Omega_{\rm PBH}}\times\\
\int_{-\infty}^{+\infty}d\ln m_{\rm H}\bigg(\frac{M_{\rm eq}}{m_{\rm H}}\bigg)^{1/2} \bar{\beta}_{m_{\rm PBH}},
\end{split}
\end{equation}
where $\boldsymbol p_{\rm mf}$ represents the parameters from the power spectrum of primordial curvature perturbation $P_{\zeta}(q,\boldsymbol p_{\rm mf})$. The corresponding total PBH abundance is defined as 
\begin{equation}\label{eq2-8}
f_{\rm PBH,th}\equiv \frac{\Omega_{\rm PBH}}{\Omega_{\rm DM}},
\end{equation}
where $\Omega_{\rm DM}$ is dark matter density parameter at present universe~\citep{CMB2018}. In order to distinguish $f_{\rm PBH}$ from the observational constraints, we label the $f_{\rm PBH}$ obtained from the power spectrum of primordial curvature perturbation is written as $f_{\rm PBH,th}$ in Eq.~(\ref{eq2-8}).

\section{Constraints on $f_{\rm PBH}$ from the lensing effect}\label{sec3}
For a lensing system, Einstein radius is one of the characteristic parameters and, taking the intervening lens with mass $m$ as a point mass, it is given by
\begin{equation}\label{eq3-1}
\theta_{\rm E}=2\sqrt{\frac{mD_{\rm LS}}{D_{\rm L}D_{\rm S}}},
\end{equation}
where $D_{\rm S}$, $D_{\rm L}$ and $D_{\rm LS}$ represent the angular diameter distance to the source, to the lens, and between the source and the lens, respectively. The lensing cross section due to a PBH lens is given by an annulus between the maximum and minimum impact parameters ($y\equiv\beta/\theta_{\rm E}$, $\beta$ stands for the source angular position),
\begin{equation}\label{eq3-2}
\begin{split}
\sigma(m, z_{\rm L}, z_{\rm S})=\pi\theta_{\rm E}^2D_{\rm L}^2(y^2_{\max}-y^2_{\min})=\\
\frac{4\pi mD_{\rm L}D_{\rm LS}}{D_{\rm S}}(y^2_{\max}-y^2_{\min}).
\end{split}
\end{equation}
It is worth emphasizing that the maximum impact parameter $y_{\max}$ and minimum impact parameter $y_{\min}$ generally depends on the observing instruments or the nature of the lensing source. For example, the maximum impact parameter $y_{\rm max}$ and minimum impact parameter $y_{\rm min}$ of FRBs micro-lensing system are determined by the maximum flux ratio of two lensed peaks and the width of signals, respectively.  For a single source, the optical depth for lensing due to a single PBH is
\begin{equation}\label{eq3-3}
\begin{split}
\tau(m,f_{\rm PBH,obs},z_{\rm S})=\int_0^{z_{\rm S}}d\chi(z_{\rm L})(1+z_{\rm L})^2\times\\
n_{\rm L}(f_{\rm PBH,obs}, m)\sigma(m,z_{\rm L},z_{\rm S})=
\frac{3}{2}f_{\rm PBH,obs}\Omega_{\rm DM}\times\\
\int_0^{z_{\rm S}}dz_{\rm L}\frac{H_0^2}{H(z_{\rm L})}
\frac{D_{\rm L}D_{\rm LS}}{D_{\rm S}}(1+z_{\rm L})^2(y^2_{\max}-y^2_{\min}),
\end{split}
\end{equation}
where $H(z_{\rm L})$ is the Hubble expansion rate at $z_{\rm L}$, $H_0$ is the Hubble constant, and $n_{\rm L}(f_{\rm PBH,obs}, m)$ is the comoving number density of the PBHs with the monochromatic mass distribution (MMD)
\begin{equation}\label{eq3-4}
n_{\rm L}(f_{\rm PBH,obs},m)=\frac{f_{\rm PBH,obs}\Omega_{\rm DM}\rho_{\rm c}}{m},
\end{equation}
where $\rho_{\rm c}$ is critical density of universe. Correspondingly, the $f_{\rm PBH}$ obtained from the lensing effect is written as $f_{\rm PBH,obs}$ in Eq.~(\ref{eq3-4}). According to the Poisson law, the probability for the null detection of lensed event is 
\begin{equation}\label{eq3-5}
P_i=\exp(-\tau_i(m,f_{\rm PBH,obs},z_{\rm S})).
\end{equation}
If we have detected a large number of astrophysical events $N_{\rm tot}$, and none of them has been lensed, the total probability of unlensed event would be given by
\begin{equation}\label{eq3-6}
P_{\rm tot}=\exp\bigg(-\sum^{N_{\rm tot}}_{i=1}\tau_i\bigg).
\end{equation}
If none lensed detection is consistent with the hypothesis that the universe is filled with the PBHs to a fraction $f_{\rm PBH,obs}$ at $100\Pi\%$ confidence level, the following condition must be valid
\begin{equation}\label{eq3-7}
P_{\rm tot}(f_{\rm PBH,obs})\geq1-\Pi.
\end{equation}
For a null search of lensed signals, then the constraint on the upper limit of $f_{\rm PBH,obs}$ can be estimated from Eq.~(\ref{eq3-7}). For the optical depth $\tau_i\ll1$, we can obtain the expected number of lensed events
\begin{equation}\label{eq3-8}
N_{\rm lensed}(m, f_{\rm PBH,obs})=\sum^{N_{\rm tot}}_{i=1}(1-\exp(-\tau_i))
\approx\sum^{N_{\rm tot}}_{i=1}\tau_i.
\end{equation}

It should be pointed out that the above formalism is only valid for the simple but widely used MMD, 
\begin{equation}\label{eq3-9}
\psi(m_{\rm PBH},m)=\delta(m_{\rm PBH}-m),
\end{equation}
where $\delta(m_{\rm PBH}-m)$ represents the $\delta$-function at the mass $m$. In fact, there is a specific EMD which corresponds to different the power spectrum of primordial curvature perturbation from different inflation models. Therefore, it is important and necessary to derive constraints on PBH with some theoretically motivated EMDs, which are closely related to realistic formation mechanisms of PBHs. For the above-mentioned EMDs, the lensing optical depth for a given event can be written as,
\begin{equation}\label{eq3-10}
\begin{split}
\tau(f_{\rm PBH,obs},z_{\rm S}, \boldsymbol p_{\rm mf})=\int dm_{\rm PBH}\int_0^{z_{\rm S}}d\chi(z_{\rm L})(1+z_{\rm L})^2 \times\\
\frac{d n_{\rm L}(f_{\rm PBH,obs},m_{\rm PBH},\boldsymbol p_{\rm mf})}{dm_{\rm PBH}}\sigma(m_{\rm PBH},z_{\rm L},z_{\rm S}),
\end{split}
\end{equation}
where $\frac{d n_{\rm L}(f_{\rm PBH,obs},m_{\rm PBH},\boldsymbol p_{\rm mf})}{dm_{\rm PBH}}$ is the comoving number density of the PBHs at EMD $\psi(m_{\rm PBH}, \boldsymbol p_{\rm mf})$
\begin{equation}\label{eq3-11}
\begin{split}
\frac{d n_{\rm L}(f_{\rm PBH,obs},m_{\rm PBH},\boldsymbol p_{\rm mf})}{dm_{\rm PBH}}=\psi(m_{\rm PBH}, \boldsymbol p_{\rm mf})\times\\
\frac{f_{\rm PBH,obs}\Omega_{\rm DM}\rho_{\rm c}}{m_{\rm PBH}}.
\end{split}
\end{equation}

Then we can derive a universal formula for connecting the constraints on the $f_{\rm PBH,obs}$ for applying constraints with the MMD to EMD. Firstly, we must respectively note the $f_{\rm PBH,obs}$ under MMD and EMD as $f_{\rm PBH,obs}^{\rm MMD}$  and $f_{\rm PBH,obs}^{\rm EMD}$. In addition, we can obtain the optical depth of single lensing source in MMD and EMD framework and note them as
\begin{equation}\label{eq3-12}
\left\{
\begin{aligned}
\tau^{\rm MMD}(m,f_{\rm PBH,obs}^{\rm MMD})=f_{\rm PBH,obs}^{\rm MMD}\tau^{\rm MMD}(m,f_{\rm PBH,obs}^{\rm MMD}=1),\\
\tau^{\rm EMD}(\boldsymbol p_{\rm mf},f_{\rm PBH,obs}^{\rm EMD})=f_{\rm PBH,obs}^{\rm EMD}\tau^{\rm EMD}(\boldsymbol p_{\rm mf},f_{\rm PBH,obs}^{\rm EMD}=1).
\end{aligned}
\right.
\end{equation}
From the relationship of the optical depth of MMD and EMD  (see Eq.~(\ref{eq3-10}) and Eq.~(\ref{eq3-3})), we can obtain that
\begin{equation}\label{eq3-13}
\begin{split}
\tau^{\rm EMD}(\boldsymbol p_{\rm mf},f_{\rm PBH,obs}^{\rm EMD})=\int_0^{+\infty} dm\psi(m_{\rm PBH},\boldsymbol p_{\rm mf})\times\\
\tau^{\rm MMD}(m_{\rm PBH},f_{\rm PBH,obs}^{\rm EMD}).
\end{split}
\end{equation}
In addition, we can obtain the upper limits of $f_{\rm PBH,obs}^{\max}$ from Eq.~(\ref{eq3-7}) and Eq.~(\ref{eq3-12}) in the MMD and EMD framework as
\begin{equation}\label{eq3-14}
\left\{
\begin{aligned}
f_{\rm PBH,obs}^{\rm MMD,\max}(m)=\frac{-\ln(1-\Pi)}{\sum^{N_{\rm tot}}_{i=1}\tau^{\rm MMD}_{i}(m,f_{\rm PBH,obs}^{\rm MMD}=1)}, \\
f_{\rm PBH,obs}^{\rm EMD,\max}(\boldsymbol p_{\rm mf})=\frac{-\ln(1-\Pi)}{\sum^{N_{\rm tot}}_{i=1}\tau^{\rm EMD}_{i}(\boldsymbol p_{\rm mf},f_{\rm PBH,obs}^{\rm EMD}=1)}.
\end{aligned}
\right.
\end{equation}
Then, we can obtain this relationship by combining Eqs.~(\ref{eq3-13},~\ref{eq3-14})
\begin{scriptsize}
\begin{equation*}\label{eq3-15}
\begin{split}
\frac{f_{\rm PBH,obs}^{\rm EMD,\max}(\boldsymbol p_{\rm mf})}{f_{\rm PBH,obs}^{\rm MMD,\max}(m)}=\frac{\sum^{N_{\rm tot}}_{i=1}\tau^{\rm MMD}_{i}(m,f_{\rm PBH,obs}^{\rm MMD}=1) }{\sum^{N_{\rm tot}}_{i=1}\tau^{\rm EMD}_{i}(\boldsymbol p_{\rm mf},f_{\rm PBH,obs}^{\rm EMD}=1)}=\\
\frac{\sum^{N_{\rm tot}}_{i=1}\tau^{\rm MMD}_{i}(m,f_{\rm PBH,obs}^{\rm MMD}=1)}{\sum^{N_{\rm tot}}_{i=1}\int_0^{\infty} dm_{\rm PBH}\tau^{\rm MMD}_{i}(m_{\rm PBH},f_{\rm PBH,obs}^{\rm EMD}=1)\psi(m_{\rm PBH},\boldsymbol p_{\rm mf})}.
\end{split}
\end{equation*}
\end{scriptsize}
Finally, we can integrate Eq.~(\ref{eq3-15}) with the same mass distribution $\psi(m,\boldsymbol p_{\rm mf})$ over $m$ to obtain that
\begin{equation}\label{eq3-16}
\begin{split}
\int_0^{\infty} dm \frac{f_{\rm PBH,obs}^{\rm EMD,\max}(\boldsymbol p_{\rm mf})\psi(m,\boldsymbol p_{\rm mf})}{f_{\rm PBH,obs}^{\rm MMD,\max}(m)}=\\
\frac{\int_0^{\infty} dm \sum^{N_{\rm tot}}_{i=1}\tau^{\rm MMD}_{i}(m,f_{\rm PBH,obs}^{\rm MMD}=1)\psi(m,\boldsymbol p_{\rm mf})}{\sum^{N_{\rm tot}}_{i=1}\int_0^{\infty} dm\tau^{\rm MMD}_{i}(m,f_{\rm PBH,obs}^{\rm EMD}=1)\psi(m,\boldsymbol p_{\rm mf})}=1.
\end{split}
\end{equation}
This relationship indicates that constraints on the $f_{\rm PBH,obs}$ from lensing effect can be perfectly consistent with the formula for applying constraints with the MMD to specific EMD~\citep{Carr2017}. Furthermore, the same relationship in Eq.~(\ref{eq3-16}) can be derived from Eq.~(\ref{eq3-8}).

\begin{figure*}
    \centering
     \includegraphics[width=0.45\textwidth, height=0.32\textwidth]{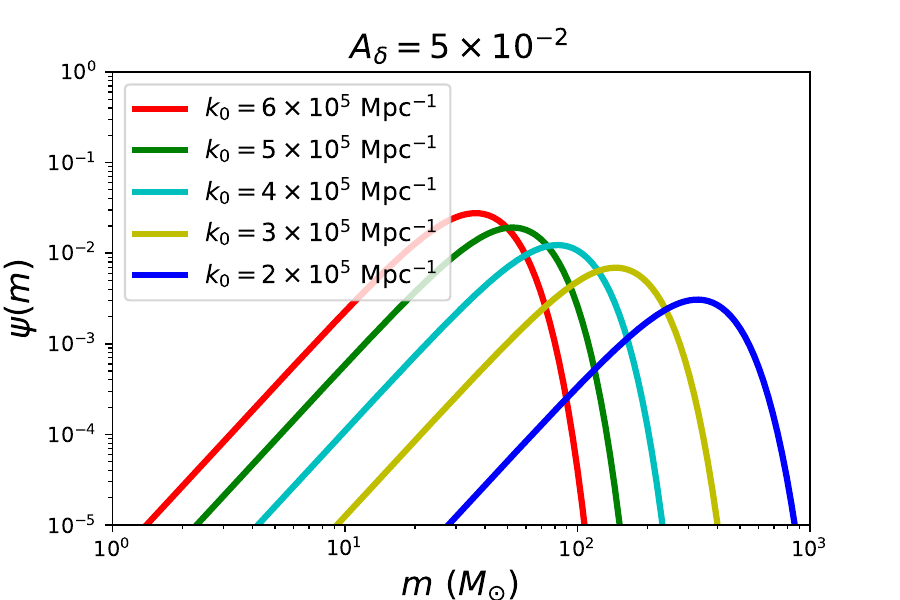}
     \includegraphics[width=0.45\textwidth, height=0.32\textwidth]{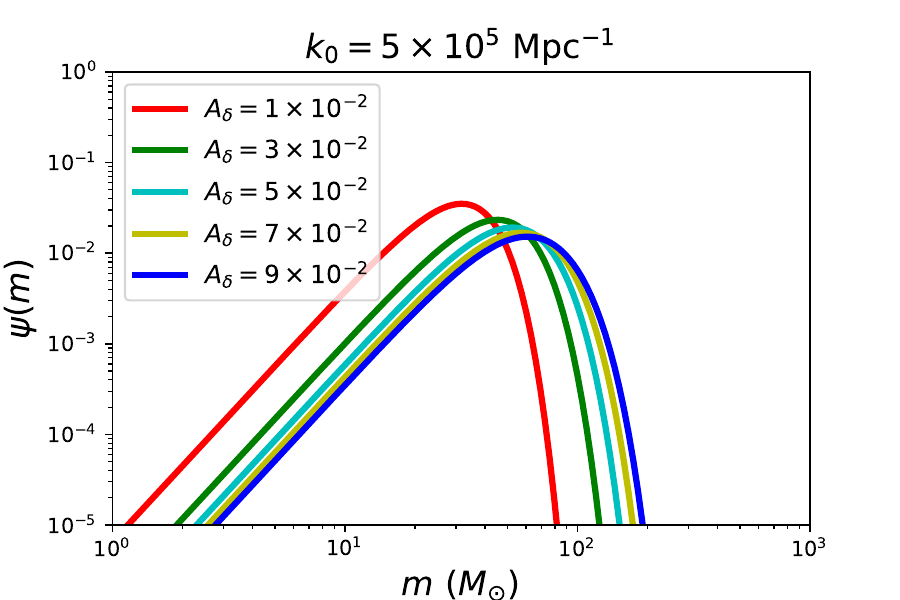}
     \caption{Mass function $\psi(m,\boldsymbol p_{\rm mf,1})$ is from $\delta$-function power spectrum of primordial curvature perturbation. {\bf Left:} Mass function $\psi(m,p_{\rm mf,1})$ correspond to the assumptions that power spectrum have fixed dimensionless amplitude $A_{\delta}=5\times10^{-2}$ and different constant wave number $k_0=[2,3,4,5,6]\times10^5~\rm Mpc^{-1}$. {\bf Right:} Mass function $\psi(m,p_{\rm mf,1})$ correspond to the assumptions that power spectrum have different dimensionless amplitude $A_{\delta}=[1,3,5,7,9]\times10^{-2}$ and fixed constant wave number $k_0=5\times10^5~\rm Mpc^{-1}$.}\label{fig1}
\end{figure*}

\section{Results}\label{sec4}
In this section, we consider two kinds of power spectrum of primordial curvature perturbation $P_{\zeta}(k,\boldsymbol p_{\rm mf})$. The first case is a $\delta$ function of $\ln k$, i.e.
\begin{equation}\label{eq4-1}
P_{\zeta}(k,\boldsymbol p_{\rm mf,1})=A_{\delta}\delta(\ln k-\ln k_0),
\end{equation}
where $\boldsymbol p_{\rm mf,1}\equiv[A_{\delta},k_0]$, $A_{\delta}$ and $k_0$ are dimensionless amplitude and constant wave number, respectively. In Fig.~\ref{fig1}, we show several examples for the mass function $\psi(m,\boldsymbol p_{\rm mf,1})$ which correspond to the $\delta$-function power spectrum of primordial curvature perturbation as Eq.~(\ref{eq4-1}). Specifically, we choose the constant wave number to be $k_0=[2,3,4,5,6]\times10^5~\rm Mpc^{-1}$ with fixed dimensionless amplitude $A_{\delta}=5\times10^{-2}$ presented in the left panel of Fig.~\ref{fig1}. Similarly, we show the mass function with different dimensionless amplitude $A_{\delta}=[1,3,5,7,9]\times10^{-2}$ and fixed constant wave number $k_0=5\times10^5~\rm Mpc^{-1}$ presented in the right panel of Fig.~\ref{fig1}.

\begin{figure*}
    \centering
     \includegraphics[width=0.45\textwidth, height=0.32\textwidth]{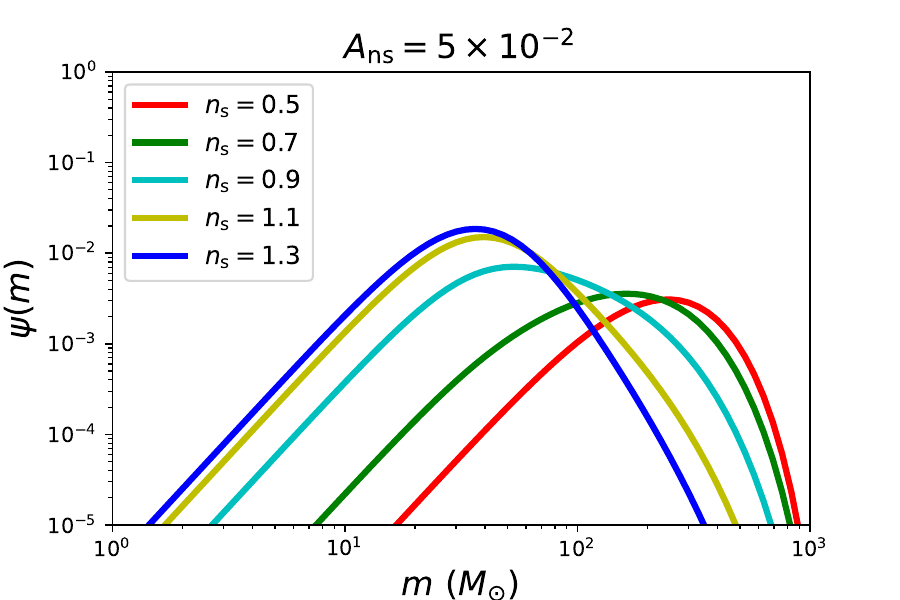}
     \includegraphics[width=0.45\textwidth, height=0.32\textwidth]{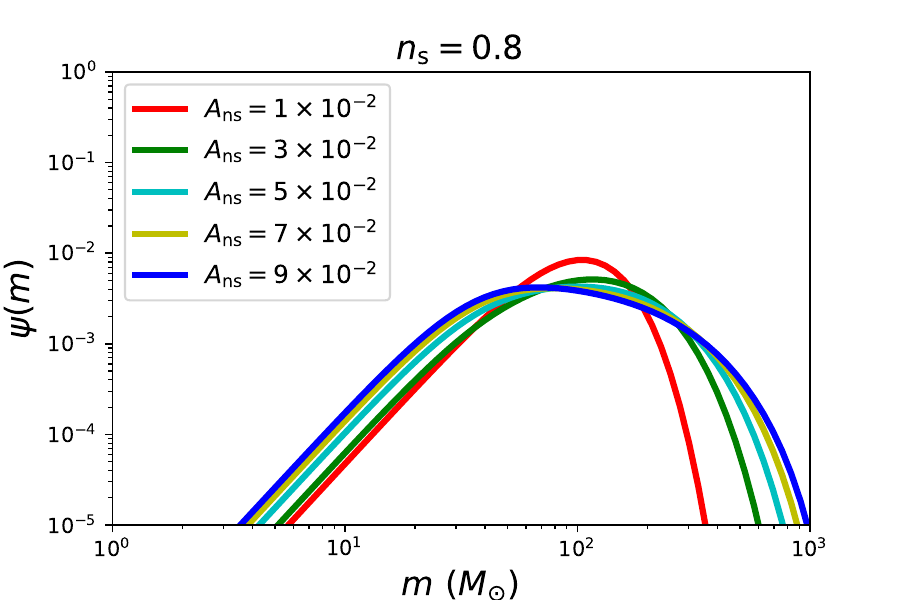}
     \caption{Same as Fig~\ref{fig1}, there is the mass function $\psi(m,\boldsymbol p_{\rm mf,2})$ which come from nearly scale invariant power spectrum of primordial curvature perturbation. {\bf Left:} Mass function $\psi(m,p_{\rm mf,2})$ correspond to the assumptions that power spectrum have fixed dimensionless amplitude $A_{\rm ns}=5\times10^{-2}$ and different spectral tilt $n_{\rm s}=[0.5,0.7,0.9,1.1,1.3]$. {\bf Right:} Mass function $\psi(m,p_{\rm mf,2})$ correspond to the assumptions that power spectrum have different dimensionless amplitude $A_{\rm ns}=[1,3,5,7,9]\times10^{-2}$ and fixed spectral tilt $n_{\rm s}=0.8$.}\label{fig2}
\end{figure*}

The second case is a nearly scale invariant shape of the form
\begin{equation}\label{eq4-1f}
\begin{split}
P_{\zeta}(k,\boldsymbol p_{\rm mf,2})=A_{\rm ns}\bigg(\frac{k}{k_{\min}}\bigg)^{n_{\rm s}-1}\times\\
\Theta(k-k_{\min})\Theta(k_{\max}-k),
\end{split}
\end{equation}
where $\boldsymbol p_{\rm mf,2}\equiv[A_{\rm ns},n_{\rm s}]$, $A_{\rm ns}$ and $n_{\rm s}$ are dimensionless amplitude and spectral tilt, respectively. In addition, We take $k_{\min}$ and $k_{\max}$ as $10^5~\rm Mpc^{-1}$ and $10^6~\rm Mpc^{-1}$, which approximately correspond to PBH mass in the range of $10~M_{\odot}$ to $10^3~M_{\odot}$. In Fig.~\ref{fig2}, we show several examples for the mass function $\psi(m,\boldsymbol p_{\rm mf,2})$ which correspond to the nearly scale invariant power spectrum of primordial curvature perturbation as Eq.~(\ref{eq4-1f}). The same as the first case, we choose the spectral tilt $n_{\rm s}=[0.5,0.7,0.9,1.1,1.3]$ with fixed dimensionless amplitude $A_{\rm ns}=5\times10^{-2}$ presented in the left panel of Fig.~\ref{fig2}. Similarly, we show the mass function with different dimensionless amplitude $A_{\rm ns}=[1,3,5,7,9]\times10^{-2}$ and fixed spectral tilt $n_{\rm s}=0.8$ presented in the right panel of Fig.~\ref{fig2}.

\begin{figure*}
    \centering
     \includegraphics[width=0.45\textwidth, height=0.33\textwidth]{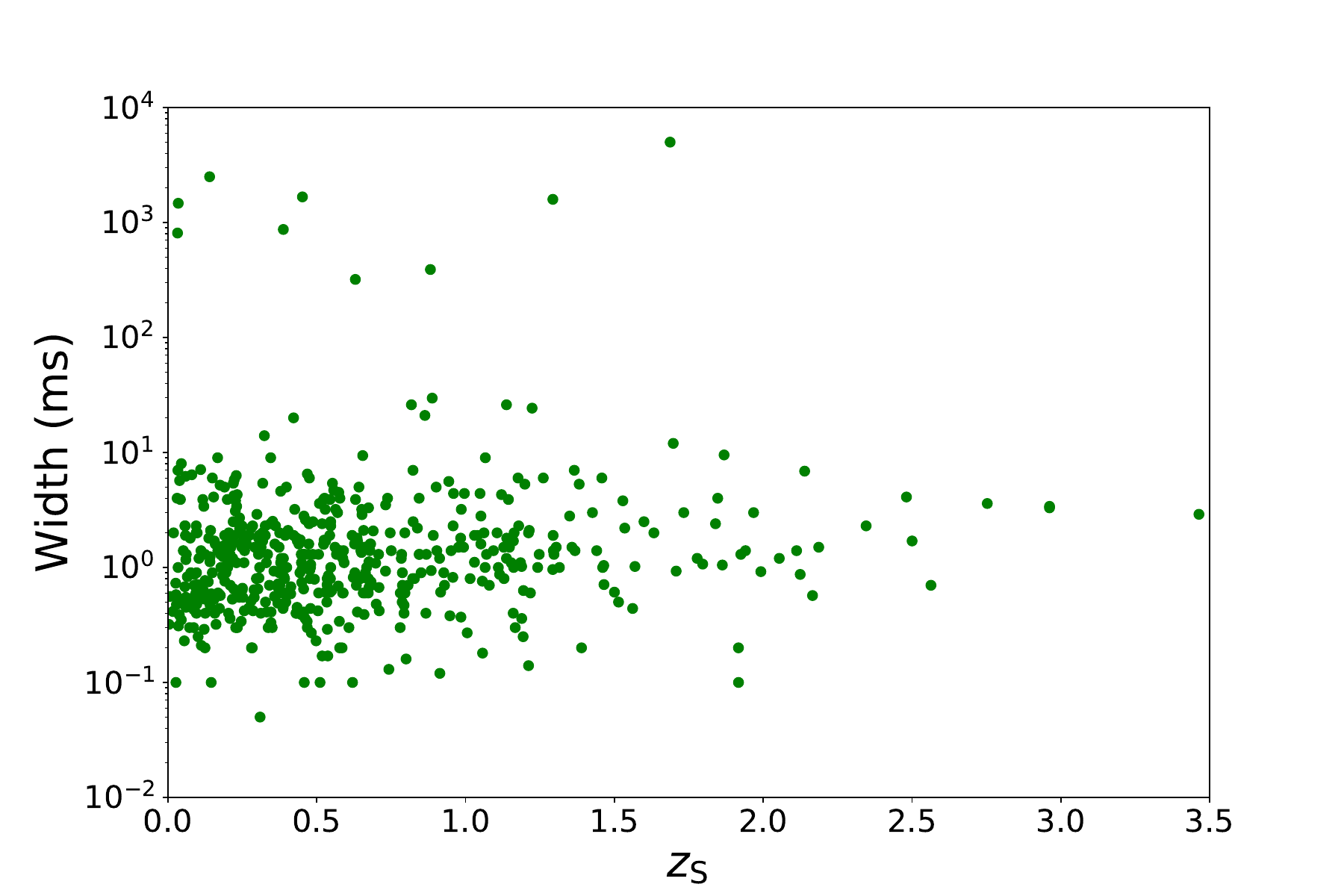}
     \includegraphics[width=0.45\textwidth, height=0.33\textwidth]{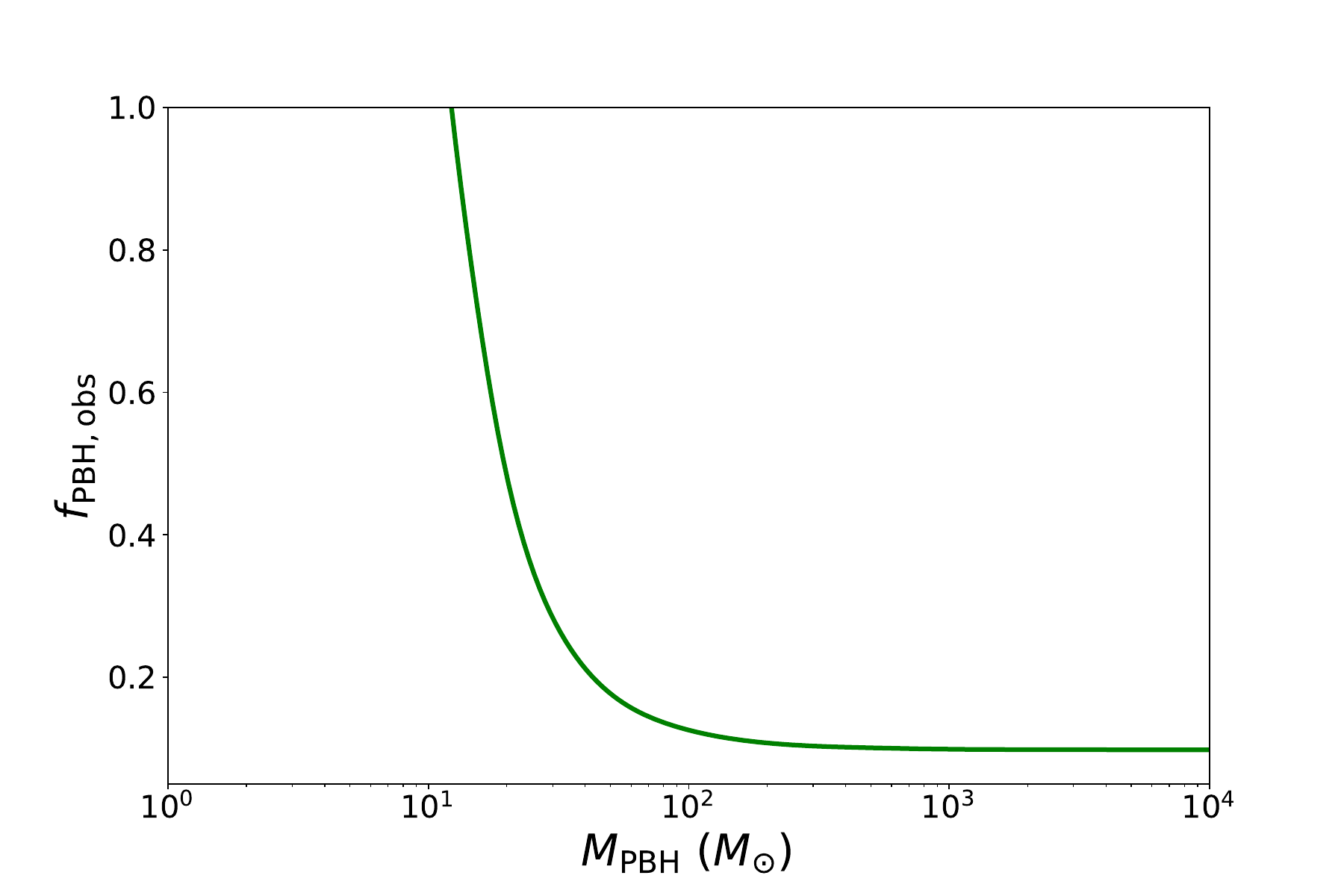}
     \caption{{\bf Left:}Two-dimensional distribution of inferred redshifts and widths for the latest 593 FRBs. {\bf Right:} Constraints on the upper limits of fraction of dark matter in the form of PBHs with the MMD from the fact that no lensed signal has been found in $593$ FRBs data.}\label{fig3}
\end{figure*}

In order to discuss the constraints on the power spectrum of primordial curvature perturbation from the lensing effect, we take FRBs as an example. At present, we use $593$ publicly available FRBs compiled by zhou et al. 2022 works~\citep{Zhou2022}. These sources consist of more than five hundred FRB events from 2018 July 25 to 2019 July 1\footnote{https://www.chime-frb.ca/catalog}~\citep{CHIME2021}. The distance and redshift of a detected FRB can be approximately estimated from its observed dispersion measure (DM), which is proportional to the number density of free electron along the line of sight and is usually decomposed into the following four ingredients,
\begin{equation}\label{eq4-2}
{\rm DM}=\frac{\rm DM_{host}+DM_{src}}{1+z}+{\rm DM_{IGM}}+{\rm DM_{MW}},
\end{equation}
where ${\rm DM_{host}}$ and ${\rm DM_{src}}$ represent DM from host galaxy and local environment, respectively. We adopt the minimum inference of redshift for all host galaxies, which corresponds to the maximum value of ${\rm DM_{host}}+{\rm DM_{src}}$ to be 200 $\rm pc/cm^{3}$. ${\rm DM_{MW}}$ is the contribution from the Milky Way. In addition, ${\rm DM_{IGM}}$ represents DM contribution from intergalactic medium (IGM). The ${\rm DM_{IGM}}-z$ relation is given by~\citep{Deng2014} and it is approximately expressed as ${\rm DM_{\rm IGM}}\sim855z~\rm pc/cm^3$ by considering the fraction $f_{\rm IGM}$ of baryon in the IGM to $f_{\rm IGM}=0.83$ and the He ionization history~\citep{Zhang2018}. DM and redshift measurements for several localized FRBs suggested that this relation is statistically favored by observations~\citep{Li2020}. We present the inferred redshifts of $593$ available FRBs in the left panel of Fig.~\ref{fig3}. For milli-lensing of FRBs, the critical value $R_{\rm f, max}$ and the width ($w$) of the observed signal determine the maximum and minimum value of impact parameter in the cross section. To ensure that both signals are detectable, the maximum value of impact parameter $y_{\max}$ can be obtained by requiring that the flux ratio of two lensed images is smaller than a critical value $R_{\rm f, max}$,
\begin{equation}\label{eq4-3}
y_{\max}=R_{\rm f,max}^{1/4}-R_{\rm f,max}^{-1/4},
\end{equation}
Here, following previous works~\cite{Munoz2016,Zhou2022,Zhou2022a,Oguri2022,Liao2021}, we take $R_{\rm f,max}=5$ for cases when we study lensing of the whole sample of all currently public FRBs. In addition,  the minimum value of impact parameter $y_{\min}$ can be obtained from the time delay between lensed signals
\begin{equation}\label{eq4-4}
\begin{split}
\Delta t=4M_{\rm PBH}\big(1+z_{\rm L}\big)\times\\
\bigg[\frac{y}{2}\sqrt{y^2+4}+\ln\bigg(\frac{\sqrt{y^2+4}+y}{\sqrt{y^2+4}-y}\bigg)\bigg]
\geq w,
\end{split}
\end{equation}
and pulse widths of all FRBs are presented in the left panel of Fig.~\ref{fig3}.

\begin{figure*}
    \centering
     \includegraphics[width=0.45\textwidth, height=0.33\textwidth]{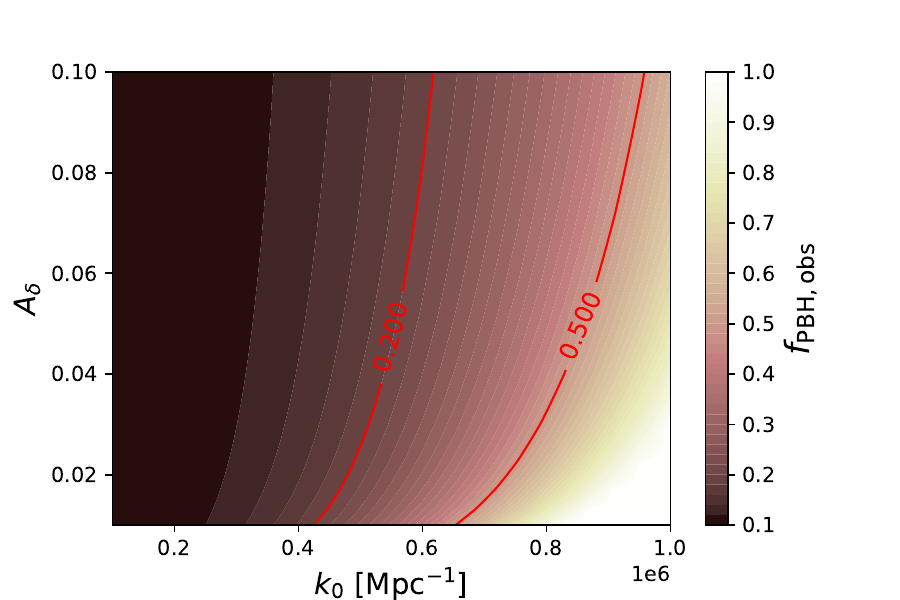}
     \includegraphics[width=0.45\textwidth, height=0.33\textwidth]{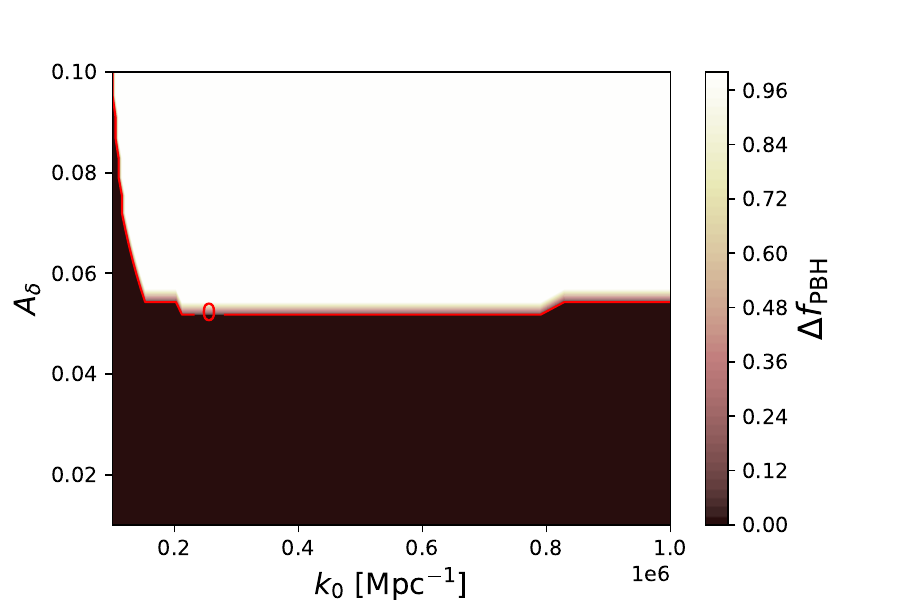}
     \caption{{\bf Left:}Constraints on the upper limits of $f_{\rm PBH,obs}$ with the first primordial curvature perturbation with two parameters ($k_{0}, A_{\delta}$) from the fact that no lensed signal has been found in $593$ FRBs data. {\bf Right:} $\Delta f_{\rm PBH}$ come from the first power spectrum of primordial curvature perturbation. The parameter space $[A_{\delta},k_0]$ can be taken in the brown area within the red solid line}\label{fig4}
\end{figure*}

\begin{figure*}
    \centering
     \includegraphics[width=0.45\textwidth, height=0.33\textwidth]{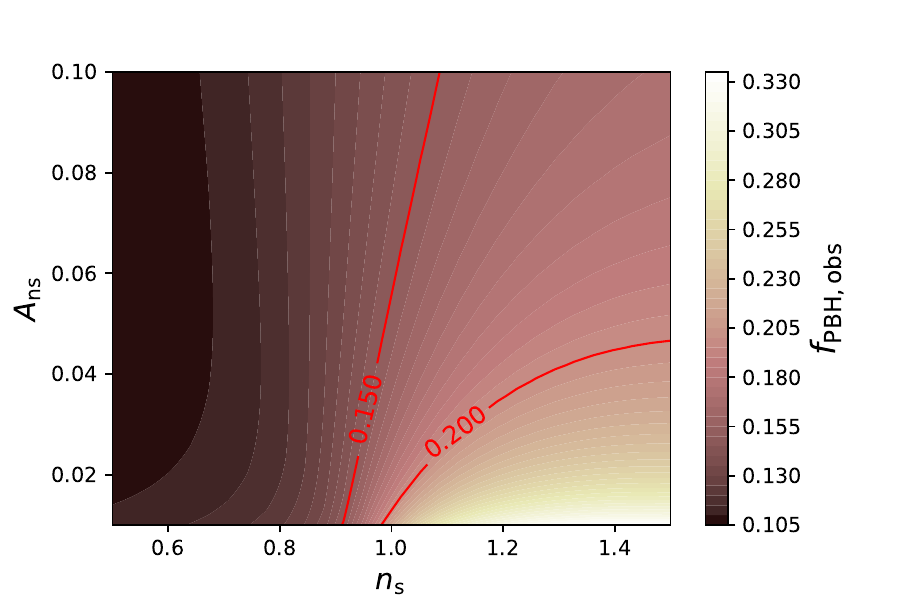}
     \includegraphics[width=0.45\textwidth, height=0.33\textwidth]{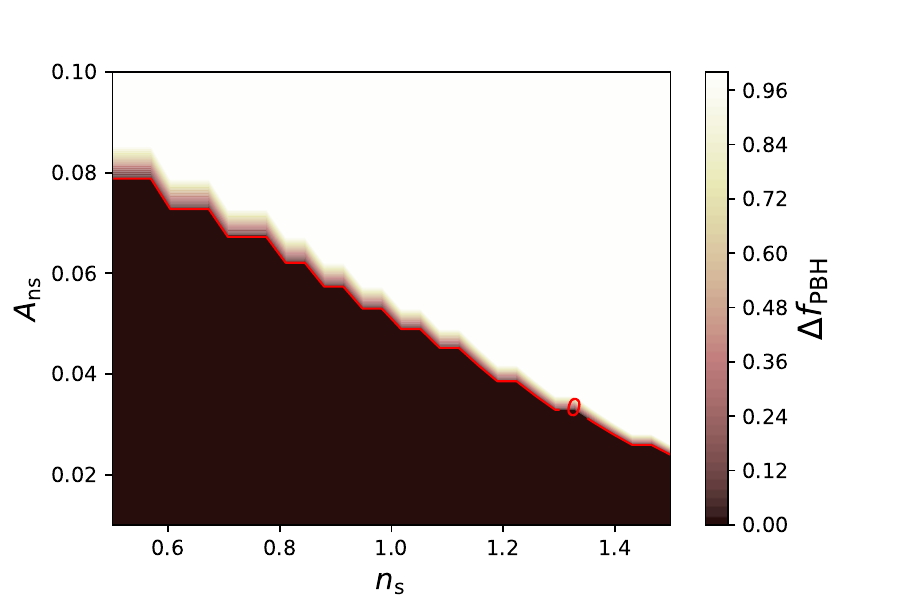}
     \caption{Same as Fig.~\ref{fig4} for the second primordial curvature perturbation with two parameters ($n_{\rm s}, A_{\rm ns}$).}\label{fig5}
\end{figure*}

After determining the the maximum and minimum value of impact parameter in the optical depth from Eqs.~(\ref{eq4-3}-\ref{eq4-4}), we can combined 593 FRBs and Eq.~(\ref{eq3-6}) to obtain the upper limit of $f_{\rm PBH,obs}$ at $100\Pi\%$ confidence level. In the right panel of Fig.~\ref{fig3}, we demonstrate the constraints on $f_{\rm PBH,obs}$ when the MMD is considered. In the $\gtrsim 10^3~M_{\odot}$ large-mass end, the constraint on $f_{\rm PBH}$ saturates to $9.8\times10^{-2}$ at 68\% confidence level. Then, From the relationship in Eq.~(\ref{eq3-16}), we derive the upper limit on $f_{\rm PBH,obs}$ corresponding to above two primordial curvature perturbation $P_{\zeta}(k,\boldsymbol p_{\rm mf})$ models, and the results are shown in the left panel of Figs.~(\ref{fig4}-\ref{fig5}). For the first primordial curvature perturbation, we vary the constant wave number $k_0$ from $10^5~\rm Mpc^{-1}$ to $10^6~\rm Mpc^{-1}$, which roughly correspond to PBHs with $\lesssim 10^3~M_{\odot}$. And, the dimensionless amplitude $A_{\delta}$ is greater than 0.01. The regions where $20\%$ and $50\%$ of $f_{\rm PBH,obs}$ have been marked by the red solid lines in the left panel of Fig.~\ref{fig4}. In addition, it should be note that the white regions in the left panel of Fig.~\ref{fig4} represent that $f_{\rm PBH,obs}$ is more than 1. In this model, less $k_0$ and larger $A_{\delta}$ corresponds to the smaller peak of the mass distribution function, which leads to improve the constraints on the $f_{\rm PBH,obs}$. In the right panel of Fig.~\ref{fig4}, the parameter space $\boldsymbol p_{\rm mf,1}\equiv[A_{\delta},k_0]$ can be allowed to exist in the brown area within the red solid line. There are two conditions under which a parameter space can be allowed to exist:
\begin{equation}\label{eq4-5}
\left\{
\begin{aligned}
f_{\rm PBH,th}\leq1,\\
\Delta f_{\rm PBH}\equiv f_{\rm PBH,th}-f_{\rm PBH,obs}\leq0.
\end{aligned}
\right.
\end{equation}
The first condition means that the density parameter of PBH from theory can not be larger than the one of dark matter. The second condition means that the theoretical prediction of PBH abundance should be lower than the upper limit of observational constraints. We find that the amplitude of primordial curvature perturbation is less than $6\times 10^{-2}$ at the scale region of $k_0\geq2\times10^5~\rm Mpc^{-1}$.

For the second case, we assume that the value of $n_{\rm s}$ varies from 0.5 to 1.5 and $A_{\rm ns}$ is greater than 0.01. The regions where $15\%$ and $20\%$ of dark matter can consist of PBHs are denoted by red solid line in the left panel of Fig.~\ref{fig5}. In this model, less $n_{\rm s}$ corresponds to the lagrer peak of the mass distribution function. In addition, larger $A_{\rm ns}$ corresponds to the broader mass distribution. The quantitative competition between the two above-mentioned effects of broadening the mass distribution, hence the constraints on the $f_{\rm PBH,th}$ is more complicated. Based on the above conditions of Eq.~(\ref{eq4-5}), we present the allowed parameter space $\boldsymbol p_{\rm mf,2}\equiv[A_{\rm ns},n_{\rm s}]$ in the brown area within the red solid line at the right panel of Fig.~\ref{fig5}. We find that the amplitude of primordial curvature perturbation is less than $5\times 10^{-2}$ at the scale invariant range ($n_{\rm s}\sim1$).

\section{Discussion}\label{sec5}
Although CMB and LSS observations have yielded strict bounds on the primordial curvature perturbation at $\rm Mpc$ scale and higher, effective constraints on primordial fluctuations on the small scales are still rare. Fortunately, the gravitational lensing effects, such as echoes of transient sources, can be used as powerful probes to constrain PBHs. Therefore, we proposed using lensing effect to constrain the primordial curvature perturbation on small scales. In this paper, we first derive the relationship that connects constraints of $f_{\rm PBH,obs}$ from the MMD to EMD for all lensing effect. Then, taking FRB as an example, we propose that its lensing effect can be used to exploring the primordial curvature perturbation. By combining 593 FRB samples~\citep{Zhou2022} and two kinds of primordial curvature perturbation models, we present the constraints on $f_{\rm PBH,obs}$ and the allowed regions of parameter space of primordial curvature perturbations in Figs.~(\ref{fig4}-\ref{fig5}). In general, null search result of lensed FRB in the latest 593 events would constrain the amplitude of primordial curvature perturbation to be less than $8\times 10^{-2}$ at the scale region of $10^5-10^6~\rm Mpc^{-1}$. Moreover, there are two significant aspects in our analysis:  
\begin{itemize}
\item When comparing the abundance of PBHs calculated by any theoretical model with the observational constraints, we should transform the results from  MMD to the EMD under the corresponding theoretical framework. For observational constraints from the lensing effect, we can use Eqs.~(\ref{eq3-16}) to translate MMD and EMD results.

\item Since the primordial power spectrum determines the mass distribution ($\psi(m,\boldsymbol p_{\rm mf})$) and theoretical abundance of PBHs ($f_{\rm PBH,th}$), it suggests that the primordial curvature perturbation parameters $\boldsymbol p_{\rm mf}$ are degenerate with the abundance of PBHs $f_{\rm PBH}$ theoretically. Therefore, if there is a tension between the predicted range of $f_{\rm PBH,obs}$ and $f_{\rm PBH,th}$ from the future lensing signals, we should consider the following possible reasons: 1. Whether the primordial perturbation model is correct; 2. Whether there are other compact dark matter, such as axion mini-clusters~\citep{Hardy2017} and compact mini halos~\citep{Ricotti2009}, participating in the observation process; 3. Whether PBHs exist evolutionary processes, such as accretion~\citep{Ricotti2007} and halo structure~\citep{Delos2023}, to change the theoretical $f_{\rm PBH,th}$ or observed physical processes.
\end{itemize}

There are several factors contributing to the uncertainties in our analysis. For example, the values for $\delta_{\rm th}$ depends on the profile of perturbations, the threshold value of the comoving density could vary from 0.2 to 0.6~\citep{Musco2013,Harada2013,Yoo2018,Musco2021}. Moreover, non-Gaussian due to the nonlinear relationship between curvature and density perturbations would lead to the amplitude of the power spectrum of primordial curvature perturbation $P_{\zeta}(q,\boldsymbol p_{\rm mf})$ might be a factor of $\mathcal{O}(2)$ larger than if we assumed a linear relationship between $\zeta$ and $\delta$~\citep{Gow2020,Luca2019,Young2019}. Finally, our analysis are based on the Press-Schechter theory. It should be noted that the statistical methods, e.g., Press-Schechter or peaks theory, would slightly affect the results~\citep{Gow2020}. It is foreseen that these constraints will be of great importance for exploring PBHs with their formation mechanisms relating to the physics of the early universe.

\section{Acknowledgements}
This work was supported by the National Key Research and Development Program of China Grant No. 2021YFC2203001; National Natural Science Foundation of China under Grants Nos.11920101003, 12021003, 11633001, 12322301, and 12275021; the Strategic Priority Research Program of the Chinese Academy of Sciences, Grant Nos. XDB2300000 and the Interdiscipline Research Funds of Beijing Normal University. H.Z is supported by China National Postdoctoral Program for Innovative Talents under Grant No.BX20230271.

\end{document}